\documentclass[reprint, amsmath,amssymb,aps,prb, superscriptaddress]{revtex4-2}
\usepackage{graphicx}
\usepackage{dcolumn}
\usepackage{bm}

\usepackage{xcolor}
\usepackage{longtable}
\usepackage{colortbl}%
  
\usepackage{booktabs,tabularx}
\usepackage{dcolumn}
\usepackage{hyperref}

\hypersetup{
    colorlinks = true,
    citecolor = {blue},
    linkcolor = {blue},
    urlcolor  = {blue},
}

\usepackage{array}
\newcolumntype{P}[1]{>{\centering\arraybackslash}p{#1}}
\newcolumntype{M}[1]{>{\centering\arraybackslash}m{#1}}

\begin{document}

\ 

\ 

\ 

\ 

\ 

\ 

\ 

\ 

\ 

\ 

\ 

\ 

\ 

\ 

\ 

\ 

\ 

\ 

\ 

\ 

\ 

\ 

\

\

\title{Rotational magnetoelastic interactions in the Dzyaloshinskii-Moriya magnet Ba$_2$CuGe$_2$O$_7$}

\author{J. Sourd}
\affiliation{Hochfeld-Magnetlabor Dresden (HLD-EMFL) and Würzburg-Dresden Cluster of Excellence ct.qmat,
Helmholtz-Zentrum Dresden-Rossendorf, 01328 Dresden, Germany}
\author{T. Kotte}
\affiliation{Hochfeld-Magnetlabor Dresden (HLD-EMFL) and Würzburg-Dresden Cluster of Excellence ct.qmat,
Helmholtz-Zentrum Dresden-Rossendorf, 01328 Dresden, Germany}
\author{P. Wild}
\affiliation{Heinz Maier-Leibnitz Zentrum (MLZ), Technische Universität München, 85748 Garching, Germany}
\author{S. Mühlbauer}
\affiliation{Heinz Maier-Leibnitz Zentrum (MLZ), Technische Universität München, 85748 Garching, Germany}
\author{J. Wosnitza}
\affiliation{Hochfeld-Magnetlabor Dresden (HLD-EMFL) and Würzburg-Dresden Cluster of Excellence ct.qmat,
 Helmholtz-Zentrum Dresden-Rossendorf, 01328 Dresden, Germany}
\affiliation{Institut für Festkörper- und Materialphysik, TU Dresden, 01062 Dresden, Germany}
\author{S. Zherlitsyn}
\affiliation{Hochfeld-Magnetlabor Dresden (HLD-EMFL) and Würzburg-Dresden Cluster of Excellence ct.qmat,
Helmholtz-Zentrum Dresden-Rossendorf, 01328 Dresden, Germany}

\date{\today}\begin{abstract}
We report the magnetoelastic properties of a Ba$_2$CuGe$_2$O$_7$ single crystal at low temperatures under a magnetic field applied along the crystallographic [001] axis. Our results extend to low temperature the $H-T$ phase diagram determined for this compound by neutron scattering. Furthermore, we observe that specific elastic modes are better sensitive to the various magnetic transitions. In particular, we observe an unusual coupling between the in-plane transverse acoustic mode and the cycloidal order at low field, which suggests a novel spin-strain mechanism originating from Dzyaloshinskii-Moriya interaction in this compound.
\end{abstract}

\maketitle

\section{\label{sec:level1}Introduction}

The study of multiferroic materials, where electric and magnetic polarizations coexist, has been a very fruitful topic in the last decades, with important outcomes for both fundamental research and technological developments \cite{spaldin2005renaissance,eerenstein2006multiferroic}. In particular, significant efforts have been made to identify the different mechanisms of multiferroic phenomena, starting from the microscopic constituents of a given material such as the spin, lattice, and orbital degrees of freedom. 

The microscopic origin of multiferroicity is discussed based on the coupling between the electric degrees of freedom, represented by the polarization vector $\mathbf{P}$, and the magnetic degrees of freedom, represented by the spin vector $\mathbf{S}$. Two main scenarios are usually considered to relate $\mathbf{P}$ and $\mathbf{S}$. Firstly, several intersite mechanisms such as the exchange striction \cite{arima2006collinear} or inverse Dzyaloshinskii-Moriya coupling \cite{katsura2005spin} link the electric polarization $\mathbf{P}_{ij}$ to two spins $\mathbf{S}_i$ and $\mathbf{S}_j$ at two different sites $\mathbf{r}_i$ and $\mathbf{r}_j$. These mechanisms predict specific electric polarization patterns for a broad class of spin textures with broken inversion symmetry, such as for spiral magnetic order \cite{mostovoy2006ferroelectricity}.
Secondly, the $d-p$ hybridization model relates the electric polarization  $\mathbf{P}_{i}$ to a single spin vector $\mathbf{S}_i$, and can be understood as a single-ion-anisotropy effect depending on the crystal-field environment of each magnetic site.

While having a different microscopic origin, these two mechanisms may in principle coexist in a given material. The case of Ba$_2$CuGe$_2$O$_7$ is particularly interesting in this aspect. This compound is a transparent insulator that crystallizes in an åkermanite-type structure (tetragonal non-centrosymmetric space group $P\bar{4}2_1m$), and only the copper ions are magnetically active. Each copper ion is surrounded by a distorted oxygen tetrahedron, and different CuO$_4$ tetrahedra are connected by corner-sharing GeO$_4$ tetrahedra, forming a two-dimensional square lattice of copper ions. The GeO$_4$ tetrahedra mediate magnetic interactions thanks to the superexchange mechanism. The resulting copper layers are separated by layers containing large barium ions, leading to much weaker magnetic interactions in the third direction. The crystal structure and magnetic properties of Ba$_2$CuGe$_2$O$_7$ are reviewed in Ref. \cite{zheludev1997field}. At low temperatures, a spiral magnetic order with the propagation vector $(1\pm \zeta, \pm \zeta,0), \zeta = 0.027$ is realized, which is a small deviation from the Néel $(1,0,0)$ order generated by the Dzyaloshinskii-Moriya interaction \cite{muhlbauer2011double}. Under magnetic field, a very rich phase diagram is observed \cite{zheludev1997field, zheludev1996spiral, chovan2002intermediate, muhlbauer2012phase}.

Since the low-energy physics of this compound originates from spin 1/2 Cu$^{2+}$ ions, single-ion anisotropy should be forbidden \cite{zheludev1996spiral}. However, theoretical studies indicate that an easy-plane anisotropy might be relevant to explain the phase diagram of Ba$_2$CuGe$_2$O$_7$ \cite{bogdanov2002magnetic}. The presence of this anisotropy term has been interpreted from the KSEA (Kaplan-Shekhtman-Entin-Wohlman-Aharony) mechanism \cite{zheludev1998experimental, zheludev1999magnetic}, which is a two-ion anisotropy generated by the Dzyaloshinskii-Moriya interaction. Taking into account this anisotropy, the complex phase diagram of Ba$_2$CuGe$_2$O$_7$ is well reproduced using a non-linear sigma model \cite{chovan2013field}.

The multiferroic phenomena in Ba$_2$CuGe$_2$O$_7$ are currently highly debated and offer a unique playground to explore the microscopic mechanisms for the coupling between electric and magnetic dipoles. For instance, in Ref. \cite{murakawa2012comprehensive} Murakawa et al. claim that the presence of electric polarization in the field-induced coplanar canted state of Ba$_2$CuGe$_2$O$_7$ excludes the inverse Dzyaloshinskii-Moriya mechanism. They introduced a $d-p$ hybridization model to explain the multiferroic properties of the Ba$_2$$T$Ge$_2$O$_7$ ($T$ = Co, Cu, Mn) familly. On the other side, in Ref. \cite{ono2020fingerprints} Ono et al. concluded that  $d-p$ hybridization is forbidden by the same argument as for the single-ion anisotropy. Moreover, the field-induced coplanar canted state should break additional symmetries allowing the inverse Dzyaloshinskii-Moriya mechanism to occur \cite{ono2020fingerprints}. An additional argument against the $d-p$ hybridization model came from the observation of an electric-polarization change above the saturation field \cite{kurihara2024field}. Furthermore, in Ref. \cite{kurihara2024field} Kurihara et al. proposed an alternative mechanism based on the coupling between magnetic dipoles and electric quadrupoles. 

In this paper, we present a detailed study of the magnetoelastic properties of Ba$_2$CuGe$_2$O$_7$ at low temperatures investigated by means of ultrasound measurements. This technique allows to scrutinize various acoustic modes of definite symmetry. This method is also sensitive to the quadrupole degrees of freedom \cite{luthi2007physical}. The high-temperature and high-energy properties of the optical phonons in Ba$_2$CuGe$_2$O$_7$ have been already examined by infrared spectroscopy \cite{nucara2014infrared}, Raman spectroscopy \cite{capitani2015raman}, and optical conductivity \cite{corasaniti2017electronic}. These studies showed a significant electron-phonon coupling, but as discussed in Ref \cite{corasaniti2017electronic}, the used high energy permits to probe only optical phonons associated to a single CuO$_4$ or GeO$_4$ tetrahedron, thus being insensitive to the effect of spin-spin correlations. The low-temperature properties of acoustic phonons in pulsed magnetic fields have been studied using ultrasound \cite{kurihara2024field}, showing the relevance of quadrupole degrees of freedom. However, no detailed study has been devoted to the magnetoelastic properties of Ba$_2$CuGe$_2$O$_7$ at low temperatures and in low DC magnetic fields, where a rich phase diagram has been observed in magnetization, specific heat, and neutron scattering \cite{muhlbauer2012phase}. In this work, we reveal an unusual coupling between the magnetic order and the transverse acoustic modes in Ba$_2$CuGe$_2$O$_7$, expected from an involvement of the quadrupolar degrees of freedom. Moreover, we show that this effect can be accounted for by considering the Dzyaloshinskii-Moriya interaction in the presence of strain. This leads to a specific spin-lattice coupling, having different symmetry signatures than the well-known exchange-striction mechanism.

First, we give a brief description of the experimental details in Sec. II. We present the results of our ultrasonic measurements for several acoustical modes in Sec. III and discuss these results in Sec. IV, with a comparison to the existing literature. In addition, we present a microscopic model in Sec. V, which permits to explain the observed spin-strain coupling for the in-plane transverse acoustic mode in the low-field cycloidal phase.
\section{Experimental Details}\label{sec2}

We used the same single crystal of Ba$_2$CuGe$_2$O$_7$ as previously studied in Refs. \cite{muhlbauer2011double, muhlbauer2012phase}. The crystal has size 1.0 mm $\times$ 2.0 mm $\times$ 3.0 mm, with the longest dimension of 3.0 mm along the [110] crystallographic direction for ultrasound propagation.

We applied magnetic fields up to 4 T along the [001] axis, because Ba$_2$CuGe$_2$O$_7$ showed the richest phase diagram for this field direction \cite{muhlbauer2012phase}. We performed ultrasound measurements utilizing the transmission pulse-echo technique with a phase-sensitive detection as described in Refs. \cite{luthi2007physical, hauspurg2024fractionalized}.  We set the ultrasound propagation direction $\textbf{k}$ perpendicular to the magnetic field $\textbf{k} \parallel [110]$, which corresponds to the direction of the nearest-neighbor copper ions, which dominate the magnetic interactions in this compound \cite{zheludev1999magnetic}. We attached LiNbO$_3$ transducers (36°-Y cut and 41°-X cut for longitudinal and for transverse modes, respectively) to the polished (110) surfaces of the single crystal. We studied three different directions of the sound-wave polarization, $\textbf{u} \parallel [110]$, $\textbf{u} \parallel [1\bar{1}0]$, and $\textbf{u} \parallel [001]$. In this work, we used ultrasound frequencies between 160 and 210 MHz.

\section{Experimental Results}\label{sec3}

We measured the sound-velocity and attenuation changes as a function of temperature at different magnetic fields, and as a function of magnetic field at different temperatures. Several phase transitions are observed, showing anomalies in the sound velocity and broad or sharp peaks in the sound attenuation. We do not observe conclusive evidence for hysteresis. Remarkably, the various transitions appear with larger anomalies for particular acoustic modes. We, thus, present the results separately for each mode.

\subsection{Transverse acoustic waves $\mathbf{k}\parallel[110], \mathbf{u}\parallel[1\bar{1}0]$}
\begin{figure*}
    \centering
    \includegraphics[width=\linewidth]{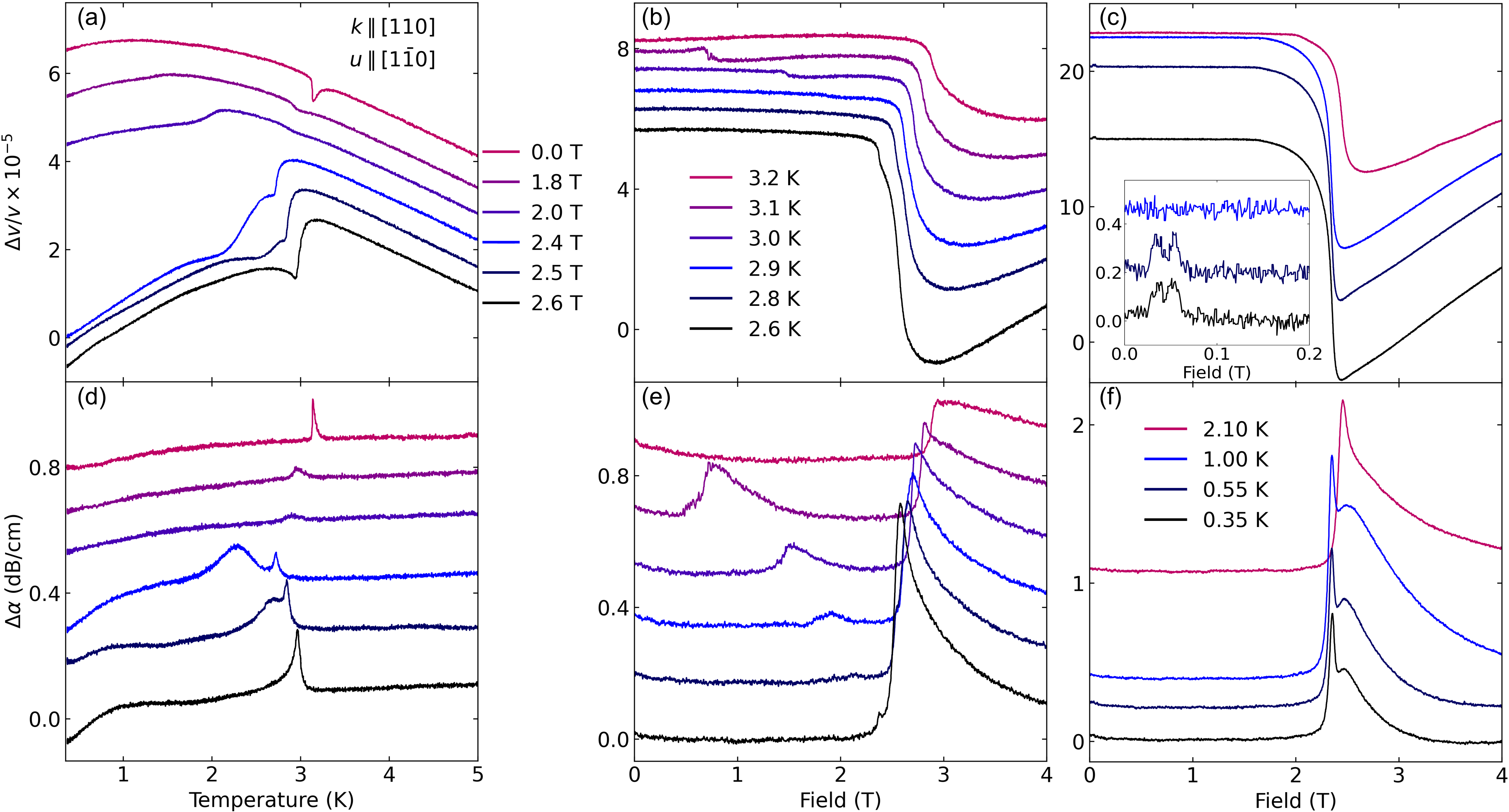}
    \caption{ (a), (b), (c) Sound-velocity and  (d), (e), (f)    sound-attenuation changes for the in-plane transverse acoustic mode $\mathbf{k}\parallel[110], \mathbf{u}\parallel[1\bar{1}0]$, versus the temperature, for selected fields, and versus the magnetic field, for selected temperatures. The curves are arbitrarily shifted for clarity. The inset of (c) shows the sound-velocity change at low fields.}\label{fig1} 
\end{figure*}
We present the temperature dependences of the sound velocity and attenuation in Fig. \hyperref[fig1]{\ref*{fig1}(a)} and Fig. \hyperref[fig1]{\ref*{fig1}(d)}, respectively. At zero field, we detect a sharp anomaly at 3.15 K for both the sound velocity and attenuation, which corresponds to the Néel temperature of Ba$_2$CuGe$_2$O$_7$ \cite{zheludev1996spiral}. With increasing field up to 2T, the anomaly is broadened and shifted towards lower temperatures, in accordance with previous specific-heat results \cite{muhlbauer2012phase}. Starting from 2 T [Fig. \hyperref[fig1]{\ref*{fig1} (a)}], we observe a second anomaly at about 2 K in the sound velocity, but not in the attenuation. At 2.4 and 2.5 T, we observe two anomalies between 2 and 3 K, in both sound velocity and attenuation. These anomalies merge into a single sharp transition at 2.6 T.

We show the field dependencies of the acoustic properties in Fig. \hyperref[fig1]{\ref*{fig1}(b)}, \hyperref[fig1]{\ref*{fig1}(c)}, \hyperref[fig1]{\ref*{fig1}(e)} and \hyperref[fig1]{\ref*{fig1}(f)}. At 3.2 K, there is a step-like anomaly for both the sound velocity and attenuation at about 3 T. At 3.1 K, we detect a second broad transition at low fields with a sound velocity jump and an attenuation peak. This corresponds to the transition between the low-field cycloidal phase and the high-field paramagnetic phase, in accordance with the temperature sweeps. Reducing the temperature, the low-field transition is shifted towards higher fields, and, surprisingly, the magnitude of the sound-velocity and attenuation anomalies is reduced. A comparable observation has been made in the specific heat \cite{muhlbauer2012phase}, where the weight of the specific heat-anomaly associated to this transition is suppressed with magnetic field. This might suggest the possibility of a critical endpoint, as observed, for example, in the multiferroic compound BiMn$_2$O$_5$ \cite{kim2009observation}. Reducing further the temperature, the step-like sound-velocity anomaly in high field splits into two separated anomalies [Fig. \hyperref[fig1]{\ref*{fig1}(b)}], and a small additional attenuation peak appears at 2.6 K [Fig. \hyperref[fig1]{\ref*{fig1}(e)}].

\ 

At low temperature, the high-field anomaly becomes more pronounced, with an attenuation change of 1.2 dB/cm at 2.1 K [Fig. \hyperref[fig1]{\ref*{fig1}(f)}], which is five times larger than the attenuation peaks observed in the temperature dependences. The sound-velocity anomaly induced by the field is also large, reaching 18$\times 10^{-5}$ at 0.35 K. Below 2.1 K, we observe a splitting of the ultrasound attenuation peak. Finally, below 0.55 K we observe two small anomalies of the sound velocity at the very low field of 0.035 and 0.05 T [inset of Fig. \hyperref[fig1]{\ref*{fig1}(c)}].

\subsection{Transverse acoustic waves $\mathbf{k}\parallel[110], \mathbf{u}\parallel[001]$ }

\begin{figure*}
    \centering
    \includegraphics[width=\linewidth]{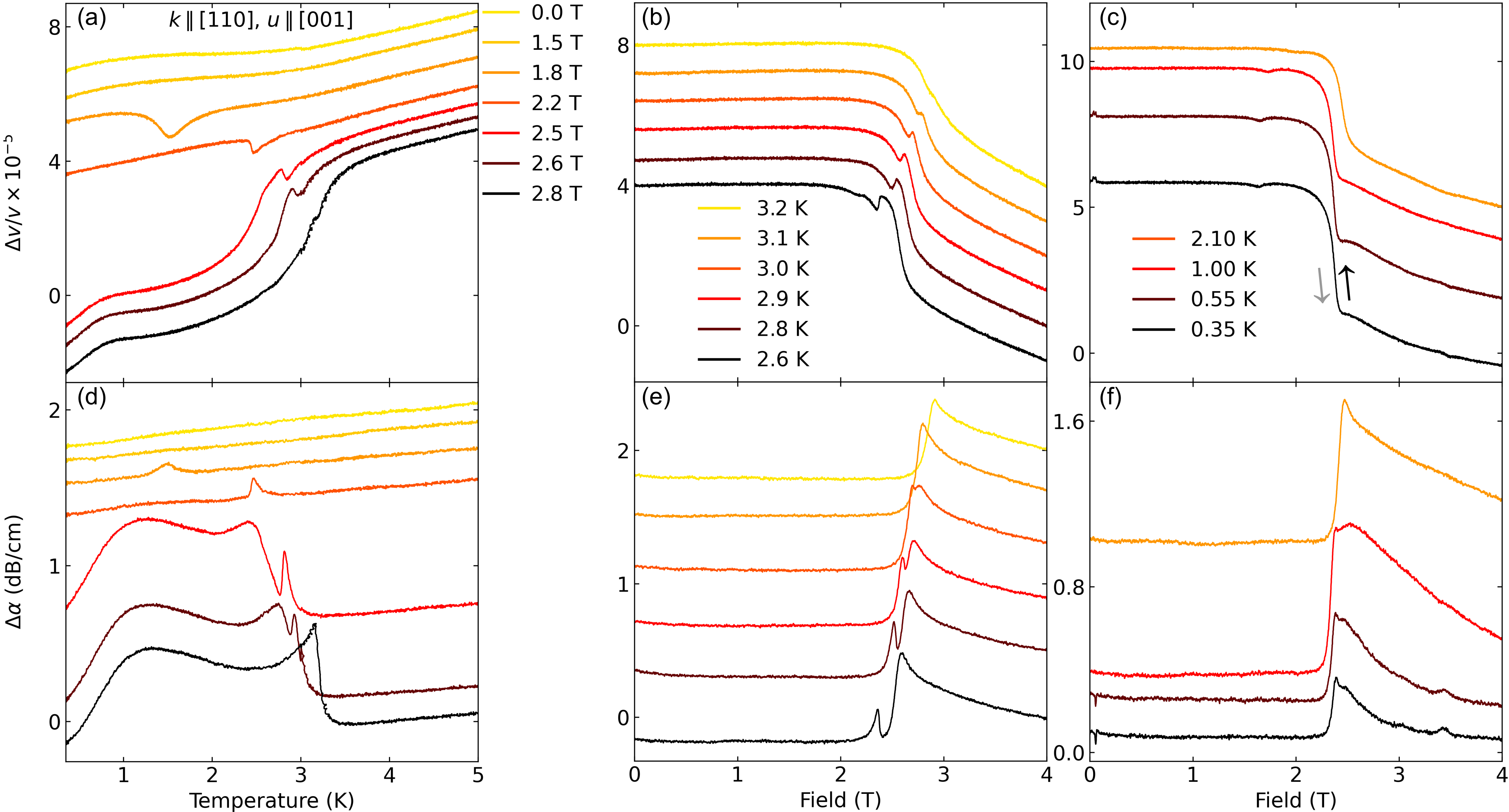}
    \caption{(a), (b), (c) Sound-velocity and  (d), (e), (f)    sound-attenuation changes for the transverse acoustic mode $\mathbf{k}\parallel[110], \mathbf{u}\parallel[001]$, versus the temperature, for selected fields, and versus the magnetic field, for selected temperatures. The curves are arbitrarily shifted for clarity.}\label{fig2} 
\end{figure*}
We present the temperature dependences of the sound velocity and attenuation in Fig. \hyperref[fig2]{\ref*{fig2}(a)} and Fig. \hyperref[fig2]{\ref*{fig2}(d)}, respectively. At zero field, there is no anomaly in the temperature dependence of the acoustic properties, in contrast to the case of the in-plane transverse mode [Fig. \hyperref[fig1]{\ref*{fig1}(a)},\hyperref[fig1]{\ref*{fig1}(d)}]. Starting from 1.8 T, we detect a pronounced dip in the sound velocity and a broad bump in the attenuation at 1.5 K, without any significant hysteresis. Thus, remarkably, the $\mathbf{u}\parallel[001]$ acoustic mode is stronger coupled to the low-temperature field-induced transition, than to the Néel transition at zero field. Moreover, the shape of the sound-velocity anomaly is different for the two transverse modes, with a pronounced jump for the $\mathbf{u}\parallel[001]$ mode for field between 2.2 and 2.6 T [Fig. \hyperref[fig2]{\ref*{fig2}(a)}].

For fields above 2.5 T, we observe an additional broad feature below about 2.5 K, with a significant phonon softening and a broad attenuation bump. This indicates stronger energy dissipation in this range of temperatures and fields. At 2.8 T, the broad bump merge with the sharp high-temperature transition.

We show the field dependencies of the acoustic properties in Fig. \hyperref[fig2]{\ref*{fig2}(b)}, \hyperref[fig2]{\ref*{fig2}(c)}, \hyperref[fig2]{\ref*{fig2}(e)} and \hyperref[fig2]{\ref*{fig2}(f)}. At 3.2 K, we observe a step-like anomaly in the sound velocity and the superposition of a step and peak in the attenuation, which are also shifted toward lower field at lower temperature. These anomalies are comparable to the one detected for the in-plane transverse mode, but twice as large. Below 3 K, the step anomaly is supplemented by a sharp jump in the sound velocity and a double peak appears in the sound attenuation [Fig. \hyperref[fig2]{\ref*{fig2}(e)}].

At 2.1 K and below, we observe a small dip at about 1.5 T and a step-like anomaly at 2.4 T in the sound velocity, and a single peak at 2.4 T in the attenuation [Fig. \hyperref[fig2]{\ref*{fig2}(f)}]. At 1 K and below, the attenuation peak splits into two, and a small additional anomaly appears at 3.4 T. Finally, we detect low-field features below 0.55 K, similar to the anomalies of the in-plane transverse acoustic mode [inset of Fig. \hyperref[fig1]{\ref*{fig1}(c)}].

\subsection{Longitudinal acoustic waves $\mathbf{k}\parallel\mathbf{u}\parallel[110]$}
\begin{figure*}
    \centering
    \includegraphics[width=\linewidth]{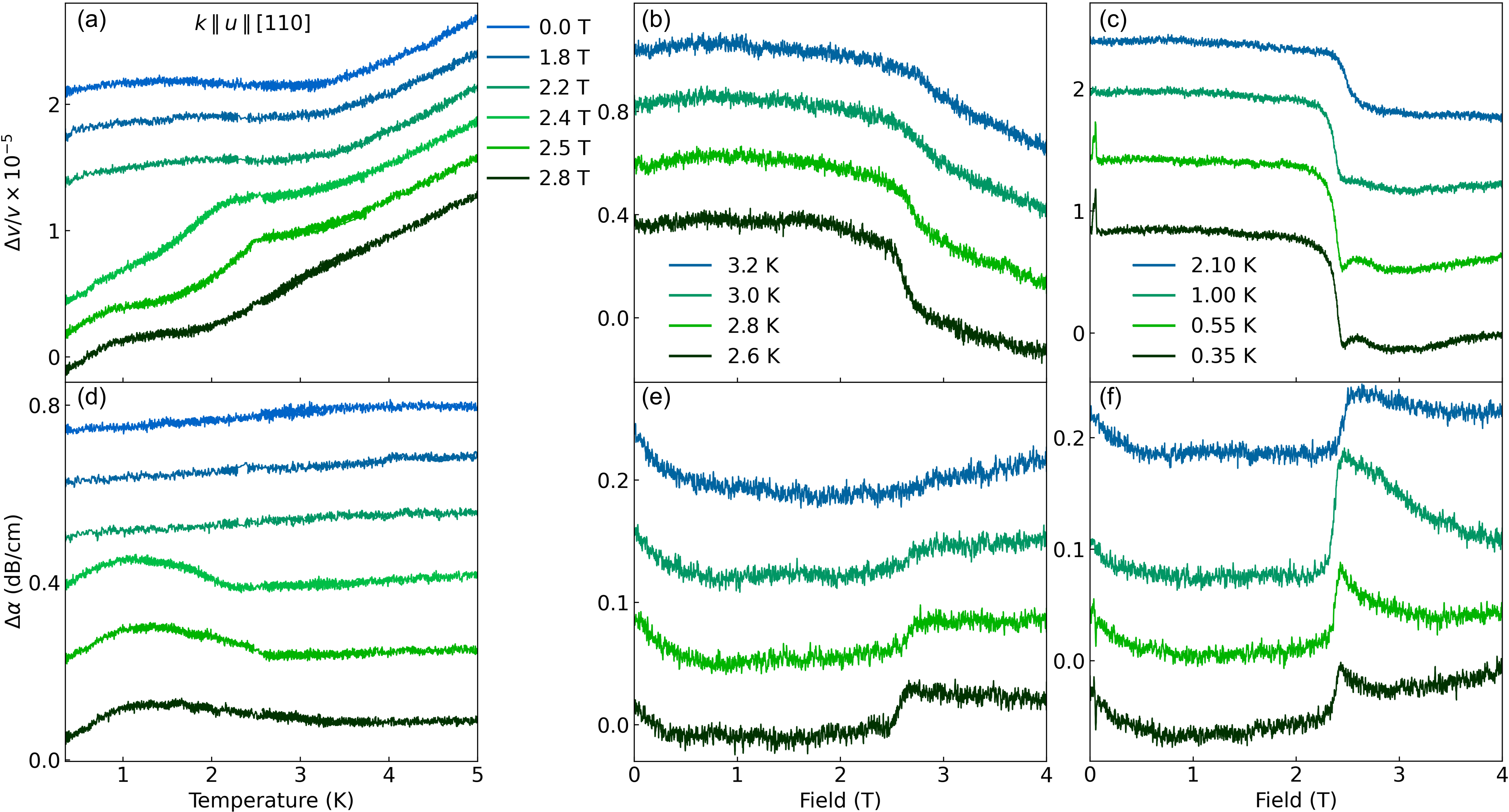}
    \caption{(a), (b), (c) Sound-velocity and  (d), (e), (f)    sound-attenuation changes for the longitudinal acoustic mode $\mathbf{k}\parallel\mathbf{u}\parallel[110]$ mode, versus the temperature, for selected fields, and versus the magnetic field, for selected temperatures. The curves are arbitrarily shifted for clarity.}\label{fig3} 
\end{figure*}
We present the temperature dependences of the sound velocity and attenuation in Fig. \hyperref[fig3]{\ref*{fig3}(a)} and Fig. \hyperref[fig3]{\ref*{fig3}(d)}, respectively. Compared to the two transverse modes the sound-velocity and attenuation changes are much less pronounced, indicating a weaker coupling between this longitudinal mode and the spin degrees of freedom. For fields below 2.4 T, we do not detect any transition. At 2.4 T, there is a pronounced kink in the sound velocity and in the attenuation at about 2.1 K, followed by a further phonon softening at lower temperature. This kink shifts toward higher temperatures with increasing magnetic field and it is suppressed at 2.8 T.

We show the field dependencies of the acoustic properties in Fig. \hyperref[fig3]{\ref*{fig3}(b)}, \hyperref[fig3]{\ref*{fig3}(c)}, \hyperref[fig3]{\ref*{fig3}(e)} and \hyperref[fig3]{\ref*{fig3}(f)}. At 2.6 K, we observe a step-like anomaly of the sound velocity and of the attenuation at 2.5 T, in agreement with our results for the other elastic modes. This step shifts towards higher fields and is suppressed with increasing temperatures.

At lower temperatures, the magnitude of the anomalies increases. Below 2.1 K, the step-like anomaly in the ultrasound attenuation transforms progressively into a peak, while a small dip appears in the sound velocity. Moreover, below 0.55 K we observe a double anomaly at 0.035 and 0.05 T in the sound velocity, similar to the anomalies of the in-plane transverse acoustic mode [inset of Fig. \hyperref[fig1]{\ref*{fig1}(c)}].

\section{Extended phase diagram and discussion}\label{sec4}

\begin{figure}
    \centering
    \includegraphics[width=\linewidth]{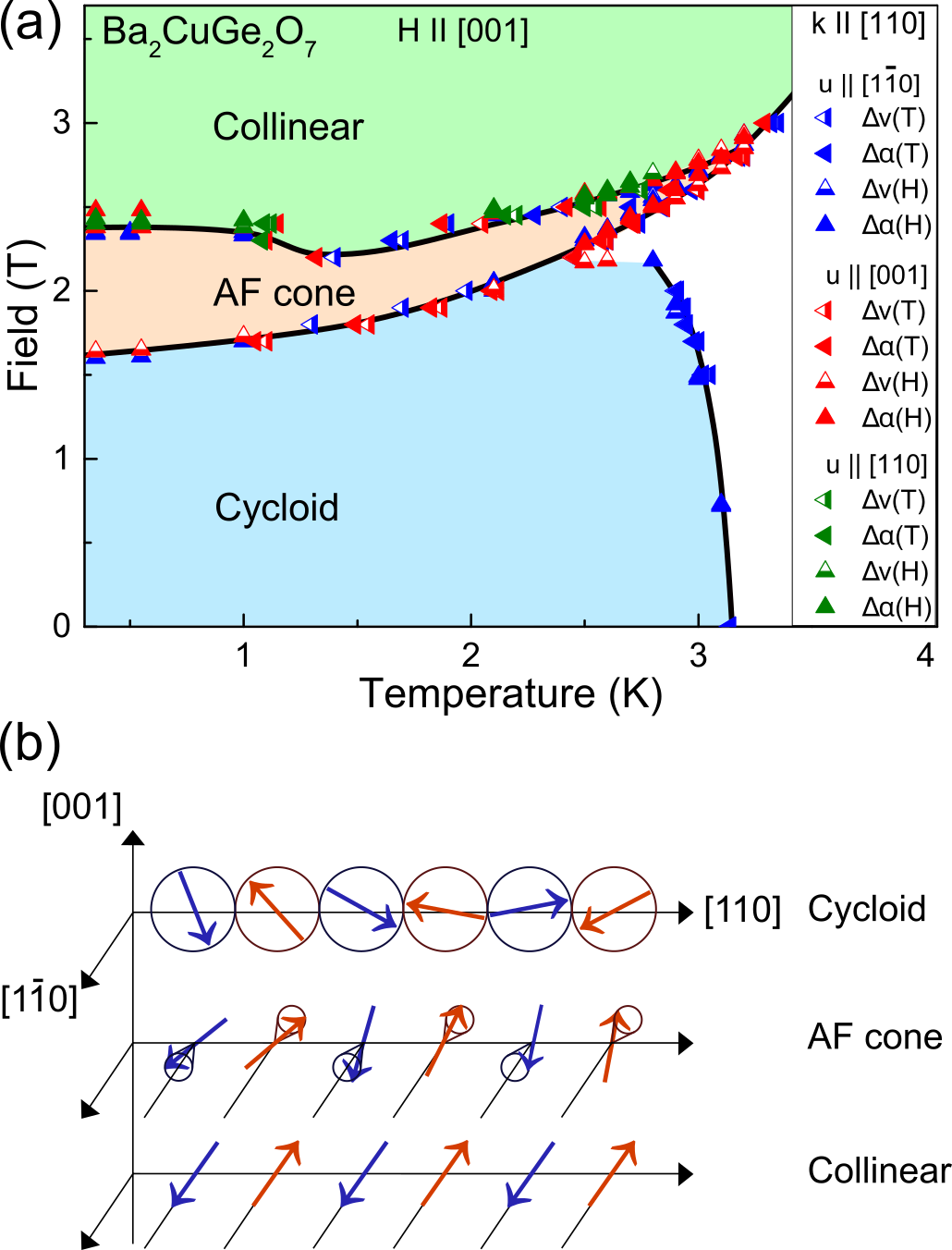}
    \caption{(a) Phase diagram constructed from the ultrasound results. (b) Sketch of the different magnetic orders determined from neutron diffraction \cite{muhlbauer2012phase}.}\label{fig4} 
\end{figure}

We summarize the detected transitions in an extended phase diagram, shown in Fig. \hyperref[fig4]{\ref*{fig4}(a)}.
Above 1.5 K, the phase diagram reproduces well the results obtained with neutron diffraction \cite{muhlbauer2012phase}, which permits us to identify the different phases observed in ultrasound. The low-field phase is associated with the soliton lattice phase, which is a distorted version of an antiferromagnetic cycloid with spiral vector $\mathbf{Q} = (1\pm \zeta, \pm \zeta,0)$. For magnetic field along the $c$ axis, both $\pm \zeta$ domains are expected to be present \cite{muhlbauer2012phase}. In this cycloidal phase, the spin direction rotates in the $\mathbf{S} \perp (1,\bar{1},0)$ plane, and the distortion increases significantly with applied magnetic field. At intermediate magnetic field, an abrupt reorientation of the magnetic moments is induced, and an AF-cone phase with a double-Q structure stabilizes \cite{muhlbauer2011double}. This cone state is comparable to some extent to the spin-flop phase of easy-axis antiferromagnets, where the dominant AF component is perpendicular to the direction of the magnetic field, and the precession plane aligns with the field \cite{muhlbauer2011double}. Finally, in high fields a commensurate phase with propagation vector $\mathbf{Q} = (1,0,0)$ is expected. These different magnetic spin arrangements are sketched in Fig. \hyperref[fig4]{\ref*{fig4}(b)}.

At the spin-flop transition of antiferromagnets, it is usual to observe step-like anomalies in the elastic properties, such as for MnF$_2$ \cite{shapira1968ultrasonic} and $\alpha-$Fe$_2$O$_3$ \cite{shapira1969ultrasonic}. Thus, our observation of field-induced step-like anomalies in Ba$_2$CuGe$_2$O$_7$ for all three elastic modes is consistent with previous observations of spin-flop transitions.

Moreover, our study shows clearly that the elastic modes couple to the different magnetic phases in a symmetry-selective fashion. The transition from the high-temperature paramagnetic phase to the antiferromagnetic cycloid is well resolved in the transverse in-plane mode $\mathbf{u} \parallel [1\bar{1}0]$ [Figs. \hyperref[fig1]{\ref*{fig1}(a)} and \hyperref[fig1]{(d)}]. In contrast, the transition into the AF-cone state is better resolved with the transverse acoustic mode $\mathbf{u} \parallel [001]$, both from the cycloidal phase upon increasing the magnetic field as well as from the paramagnetic phase upon decreasing the temperature [Fig. \hyperref[fig2]{\ref*{fig2}}]. This is particularly visible from the temperature-dependent curves at 1.8 T, where we observed a clear dip in the sound velocity and a broad attenuation bump, whereas no significant anomaly is detected for the other two modes. Finally, we clearly observed the transition into the collinear phase for all three acoustic modes studied here.

In order to get a better insight into these symmetry aspects, the comparison with the case of the spin-density wave of chromium is particularly interesting. In chromium, only the longitudinal acoustic modes show an anomaly at the Néel transition \cite{muir1987elastic}, whereas the transverse acoustic modes couple to the low-temperature spin-reorientation transition. This transition separates the transverse SDW phase at high temperature, where the spin direction $\mathbf{S}$ is perpendicular to the ordering vector $\mathbf{Q}$, and the longitudinal SDW with $\mathbf{S} \parallel \mathbf{Q}$ at low temperature \cite{fawcett1988spin}. In our case, the transition between the cycloidal phase with $\mathbf{S} \parallel \mathbf{Q}$ and the AF-cone phase with $\mathbf{S} \perp \mathbf{Q}$ is also associated with a reorientation of the spin direction and coupled to the transverse acoustic modes. However, contrary to the case of chromium, for Ba$_2$CuGe$_2$O$_7$ the longitudinal mode does not couple to the antiferromagnetic ordering at zero field.

Generally, for magnetic transitions, we expect a stronger coupling of the longitudinal acoustic modes to the order parameter \cite{luthi2007physical,tachiki1975criterion}. The reason is that longitudinal elastic waves lead to a compression (expansion) of the lattice, whereas to the transverse modes are volume conserving, as for example found in FeF$_2$ \cite{ikushima1971acoustic} and MnF$_2$ \cite{melcher1970elastic}. Ba$_2$CuGe$_2$O$_7$ is thus a clear counter example to this rule.  A possible scenario might be the presence of quadrupolar or nematic degrees of freedom associated with the Néel transition. Such quadrupole degrees of freedom should not couple strongly to longitudinal acoustic modes, which would explain our results. The authors of Ref. \cite{kurihara2024field} proposed a coupling between magnetic dipoles and electric quadrupoles to explain the properties of Ba$_2$CuGe$_2$O$_7$ above 20 T. Furthermore, the presence of spin-nematic interactions in the closely related compound Ba$_2$CoGe$_2$O$_7$ has been proposed to explain the in-plane anisotropy observed in neutron-scattering experiments \cite{soda2014spin}. Thus, quadrupole degrees of freedom might explain the low-field behavior of Ba$_2$CuGe$_2$O$_7$ as well. Another possible scenario is to consider the peculiarity of indirect-exchange interactions between copper ions, which are mediated by GeO$_4$ tetrahedra in Ba$_2$CuGe$_2$O$_7$. From the Goodenough-Kanamori rule \cite{goodenough1955theory,kanamori1959superexchange}, one expects a strong variation of the magnetic interactions upon varying the angles in the exchange paths. In Ref. \cite{wang2016field} Wange et al. suggested that such angular effect might explain the difference of the magnetoelastic coupling to longitudinal or transvserse acoustic modes in Sr$_3$Cr$_2$O$_8$. In the next section, we propose a third scenario that is based on peculiarities of the Dzyaloshinskii-Moriya interaction, and develop a toy model that is consistent with our experiments.

For the transition from the cycloidal to the AF-cone state at intermediate fields, our study shows opposite anomalies of the two transverse modes. While the magnetic field induces a hardening of the acoustic $\mathbf{u} \parallel [001]$ mode [with a jump in Fig. \hyperref[fig2]{\ref*{fig2}(b)}], it produces a softening of the in-plane acoustic mode $\mathbf{u} \parallel [1\bar{1}0]$ [Fig. \hyperref[fig1]{\ref*{fig1}(b)}]. In Refs. \cite{plumer1982magnetoelastic,nii2014elastic}, it is shown that in the cone state of MnSi, the longitudinal mode propagating along the cone opening exhibits a hardening, while the longitudinal mode perpendicular to the cone opening displays a softening. Our study shows a softening of the longitudinal  $\mathbf{u} \parallel [110]$ mode, which is in agreement with the cone opening along $[1\bar{1}0]$. In the cone state of MnSc$_2$Se$_4$, a hardening occurs for the transverse acoustic mode with polarization perpendicular to the cone opening \cite{sourd2024magnon}. Thus, our results suggest a cone opening perpendicular to [110] and [001], in line with the neutron-diffraction data \cite{muhlbauer2012phase}.

Finally, our low-temperature results indicate the presence of additional features. At very low fields, we clearly observed two small anomalies in all acoustic modes. Murakawa et al. showed that at 2 K a small magnetic field of 0.1 T permits to induce a single magnetic domain in Ba$_2$CuGe$_2$O$_7$ \cite{murakawa2009electric}. Thus, our low-field anomalies might be related to the domain structure of Ba$_2$CuGe$_2$O$_7$. Moreover, below 2.1 K and around 2.5 T, we observe in all acoustic modes a double anomaly, at the boundary between the AF-cone state and the AF collinear state. These anomalies correspond to some extent to a reverse spin-flop transition, since the collinear state is stabilized at high fields. Close to a spin-flop transition, a phase separation regime has been observed in MnF$_2$ \cite{king1973spin}. Thus, our observation of a double transition could correspond to a phase separation regime. A second possible scenario might be a two-stage spin-flop transition induced by the Dzyaloshinskii-Moriya interaction, as observed in LaCuO$_4$ \cite{thio1990magnetoresistance} and BaCu$_2$Si$_2$O$_7$\cite{tsukada2001two}.

\section{Toy model and rotational magnetoelastic coupling}\label{sec4}
\begin{figure}
    \centering
    \includegraphics[width=\linewidth]{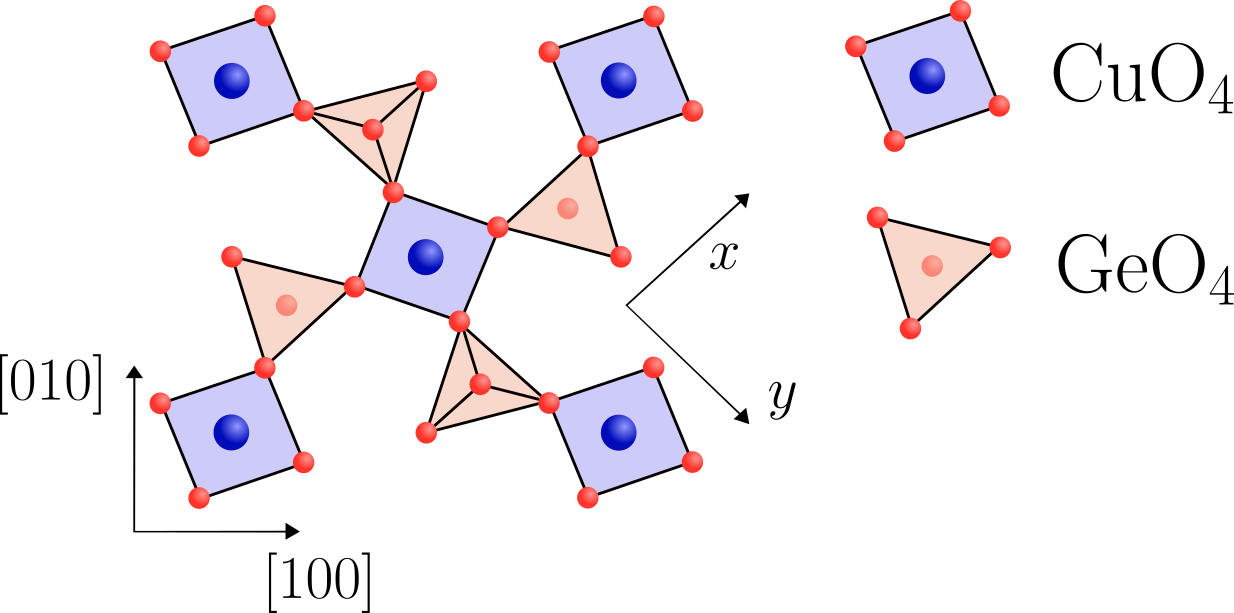}
    \caption{Crystal structure in the copper plane of Ba$_2$CuGe$_2$O$_7$.}\label{fig5} 
\end{figure}

In this section, we propose a microscopic mechanism for the spin-strain coupling that explains our ultrasound results at zero field. We start from the microscopic Hamiltonian of Ref. \cite{zheludev1999magnetic}, which considers an isolated copper plane and reproduce the spin-wave spectrum of Ba$_2$CuGe$_2$O$_7$ with nearest-neighbor exchange and Dzyaloshinskii-Moriya interactions $J$ and $D$ respectively, and an anisotropy term $D^2/2J$. We neglect the anisotropy term to simplify the discussion. Denoting the spin directions as $\alpha,\beta = x, y, z$, the spin Hamilonian of Ref. \cite{zheludev1999magnetic} is written in a matrix form as:
\begin{equation}\label{eq1}
    \begin{split}
            \mathcal{H}_s &= \sum_{i\delta\alpha\beta}S_i^\alpha \bar{J}_{\delta}^{\alpha\beta}S_{i+\delta}^\beta, 
            \\ \bar{J}_{x} &= \begin{pmatrix} J & 0 & D \\ 0 & J & 0 \\ - D & 0 & J \end{pmatrix}, 
        \ \  \bar{J}_{y} = \begin{pmatrix} J & 0 & 0 \\ 0 & J &  D \\ 0 & - D & J \end{pmatrix},
    \end{split}{}
\end{equation}{}
where $\delta$ runs over the two nearest neighbors of a given cooper atom on the square lattice in the $+\mathbf{x}$ and $+\mathbf{y}$ directions. As shown in Fig. \hyperref[fig5]{\ref*{fig5}}, this square lattice is rotated by $\pi/2$ compared to the tetragonal axis of Ba$_2$CuGe$_2$O$_7$, so that $x \parallel [110]$ and $y \parallel [1\bar{1}0]$.
We then evaluate the spin-lattice interactions following the procedure introduced in Ref. \cite{sourd2024magnon}. Atomic displacements are taken into account by writing the position of a given Cu$^{2+}$ ion as $\mathbf{R}_i = \mathbf{R}_i^0 + \mathbf{Q}_i$, with the equilibrium position $\mathbf{R}_i^0$ and the displacement $\mathbf{Q}_i$. With the bond $\boldsymbol{\delta} = \mathbf{R}_i - \mathbf{R}_{i+\delta}$, the Taylor expansion of the coupling parameters $\bar{J}_{\delta}^{\alpha\beta}$ around the equilibrium atomic positions gives:
\begin{equation}\label{eq2}
    \begin{split}
        \bar{J}_{\delta}^{\alpha\beta} \approx (\bar{J}_{\delta}^{\alpha\beta})^0 &+ \sum_{\mu} \frac{\partial \bar{J}_{\delta}^{\alpha\beta}}{\partial R^\mu}Q_\delta^{\mu} + \frac{1}{2} \sum_{\mu,\nu} \frac{\partial^2 \bar{J}_{\delta}^{\alpha\beta}}{\partial R^\mu\partial R^\nu}Q_\delta^{\mu}Q_\delta^{\nu},
    \end{split}{}
\end{equation}{}
where $Q_\delta^{\mu} = Q_i^{\mu} - Q_{i+\delta}^{\mu}$ and $\mu,\nu = x,y,z$ are Cartesian coordinates. The atomic displacement $\mathbf{Q}_i$ at the atomic position $i$ is then quantized with the usual phonon destruction and creation operators $c_{k\lambda}$ and $c_{k\lambda}^\dagger$, respectively, of momentum $k$ and polarization $\lambda$ \cite{mahan2000many}:
\begin{equation}\label{eq3}
    \begin{split}
        \mathbf{Q}_i &= i \sum_{k\lambda} \sqrt{\frac{1}{2M_0N\omega_{k\lambda}^0}} \mathbf{u}_{k\lambda} (c_{k\lambda}+c_{-k\lambda}^\dagger)e^{i\mathbf{k}\cdot\mathbf{R}_i^0},
    \end{split}{}
\end{equation}{}
where $M_0$ denotes the mass of the Cu$^{2+}$ ion, $N$ the number of copper sites,  $\mathbf{u}_{k\lambda} = -\mathbf{u}_{-k\lambda}$ the phonon polarization vector, and $\omega_{k\lambda}^0$ the energy of the phonon excitation. In the following, we will consider each phonon polarization independently, and we thus drop the index $\lambda$. Taking the dynamics of the phonons to be harmonic and using the expansion of Eq. \hyperref[eq2]{(\ref*{eq2})} in the spin Hamiltonian of Eq. \hyperref[eq1]{(\ref*{eq1})}, the lattice Hamiltonian and the spin-lattice coupling can be written as:
\begin{equation}\label{eq4}
    \begin{split}
       \mathcal{H}_{l} &= \sum_{k} \omega_{k}^0 c_{k}^\dagger c_{k},\\
            \mathcal{H}_{sl} &= \sum_k U_k^1 C_k + \frac{1}{2}\sum_{kk'} U_{kk'}^2 C_kC_{k'},
    \end{split}{}
\end{equation}{}
respectively, where $C_k = c_k + c_{-k}^\dagger$ is the phonon displacement operator. $U^1_k$ and $U^2_{kk'}$ are given by:
\begin{equation}\label{eq5}
    \begin{split}
        U_k^1 &= \frac{1}{\sqrt{2M_0N\omega_{k}^0}} \sum_{i\delta\alpha\beta} S_i^\alpha S_{i+\delta}^{\beta} \left(e^{i\mathbf{k}\cdot\mathbf{R}_i}-e^{i\mathbf{k}\cdot\mathbf{R}_{i+\delta}}\right)\\
        & \ \ \ \ \times \left(i \sum_\mu u_k^\mu \frac{\partial \bar{J}_{\delta}^{\alpha\beta}}{\partial R^\mu}\right), \\ 
        U_{kk'}^2 &=  \frac{1}{2M_0N\sqrt{\omega_{k}^0\omega_{k'}^0}}\sum_{i\delta\alpha\beta} S_i^\alpha S_{i+\delta}^{\beta} \left(e^{i\mathbf{k}\cdot\mathbf{R}_i}-e^{i\mathbf{k}\cdot\mathbf{R}_{i+\delta}}\right)\\
        & \ \ \ \ \times\left(e^{i\mathbf{k}'\cdot\mathbf{R}_i}-e^{i\mathbf{k'}\cdot\mathbf{R}_{i+\delta}}\right) \left(i^2 \sum_{\mu\nu} u_k^\mu u_{k'}^\nu \frac{\partial^2 \bar{J}_{\delta}^{\alpha\beta}}{\partial R^\mu\partial R^\nu}\right).
    \end{split}{}
\end{equation}{}
Since the $U_k^1$ term in $\mathcal{H}_{sl}$ is linear in the phonon operator, its expectation value is zero unless we consider interactions with fluctuations of the spin operators $S_i^\alpha$. However, the $U_{kk'}^2$ term is quadratic in phonon operators, and, thus generates immediately a change of the phonon energy $\Delta \omega_k^0$. In order to discuss the symmetry content of the theory, we discard the $U_k^1$ term and write the change of the sound velocity, $\frac{\Delta v}{v}$, up to a shift $\Delta_0$ as:
\begin{equation}\label{eq6}
    \begin{split}
        \left(\frac{\Delta v}{v}\right)' &= \frac{\Delta v}{v} + \Delta_0 = \lim_{k\rightarrow 0} \frac{\Delta \omega_k^0}{\omega_k^0} = \lim_{k\rightarrow 0} \frac{U_{kk}^2}{\omega_k^0}.
    \end{split}{}
\end{equation}{}
For homogeneous spin configurations, we use a locally rotating reference frame in order to write the products $S_i^\alpha S_{i+\delta}^{\beta}$ as $S_0^\alpha S_{\delta}^{\beta}$, where $\mathbf{S}_0 = (0,0,S)$. Denoting $\mathbf{e}_k$ the unit vector in the direction of $\mathbf{k} = k\mathbf{e}_k$, and using $\omega_k^0 = v k$ for the phonon energy in the long-wavelength limit, the variation of the sound velocity is evaluated as:
\begin{equation}\label{eq7}
    \begin{split}
   \left(\frac{\Delta v}{v}\right)' &=  \sum_{\delta\alpha\beta} \frac{S_0^\alpha S_\delta^\beta}{M_0 v^2} \left(\mathbf{e}_k \cdot \boldsymbol{\delta}\right)^2\left(i^2 \sum_{\mu\nu} u_k^\mu u_{k}^\nu \frac{\partial^2 \bar{J}_{\delta}^{\alpha\beta} }{\partial R^\mu\partial R^\nu}\right).
    \end{split}{}
\end{equation}{}
Thus, from the factor $\mathbf{e}_k \cdot \boldsymbol{\delta}$, the bonds perpendicular to the sound-wave propagation direction $\mathbf{k}$ will not contribute, and in order to match our experiments with $\mathbf{k}\parallel[110]$ we have only to consider the bond along $\mathbf{x}$. The key ingredient of the theory is the derivative of the exchange parameters $\partial^2 \bar{J}_{\delta}^{\alpha\beta}/\partial R^\mu\partial R^\nu$. For the exchange coupling $J = \bar{J}_{\delta}^{\alpha\alpha}$, a commonly used ansatz is an isotropic exponential dependence of the form $\bar{J}_{\delta}^{\alpha\alpha}  = J\text{exp}(-||\boldsymbol{\delta}||/\xi)$ \cite{bhattacharjee2011interplay}, where $\xi$ is the characteric length. In that case, the derivatives are easily evaluated: 
\begin{equation}\label{eq8}
    \begin{split}
        \left(i^2 \sum_{\mu\nu} u_k^\mu u_{k}^\nu \frac{\partial^2 \bar{J}_{\delta}^{\alpha\alpha} }{\partial R^\mu\partial R^\nu}\right) &= -\frac{J}{\xi^2}(\boldsymbol{\delta} \cdot \mathbf{u}_k)^2.
    \end{split}{}
\end{equation}{}
Applying this result to our toy model with propagation along $\mathbf{x}$, we obtain that only the longitudinal mode contribute, while the contribution from the two transverse modes is zero since in that case $\mathbf{u}_k \perp \mathbf{x}$. This is in clear contradiction with our experimental results. 

\begin{figure}
    \centering
    \includegraphics[width=\linewidth]{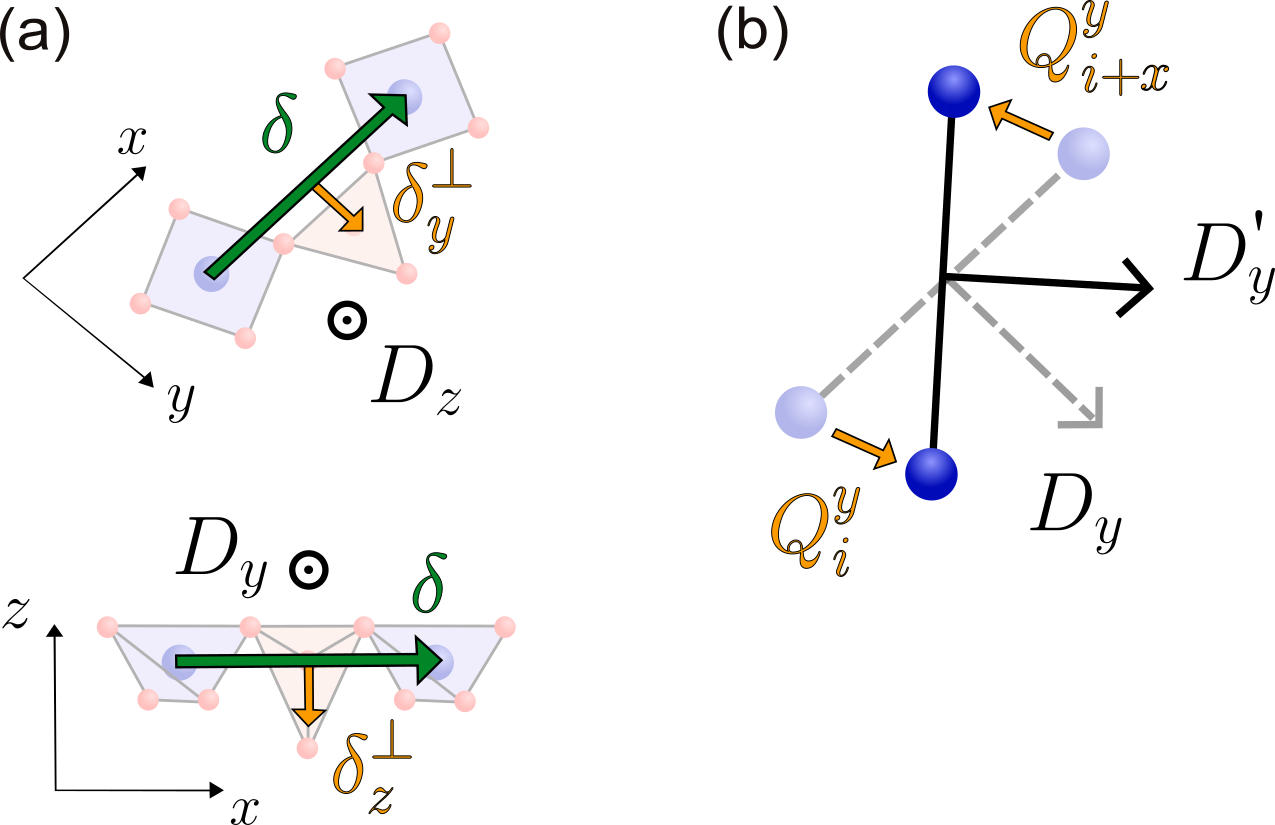}
    \caption{(a) Orientation of the Dzyaloshinskii-Moriya vector as a function of the bond vector $\boldsymbol{\delta}$ and the vector to the superexchange center $\boldsymbol{\delta}^\perp$ in the $(x,y)$ and $(x,z)$ plane. (b) Rotation of the Dzyaloshinskii-Moriya interaction vector induced by the in-plane transverse acoustic wave $\mathbf{k} \parallel [110]$, $\mathbf{u}\parallel [1\bar{1}0]$.}\label{fig6} 
\end{figure}

We thus propose an alternative mechanism for the spin-lattice coupling. The previous results show that we should take into account the variation of angles associated to microscopic displacements, rather than the variation of distances which generates a coupling with the longitudinal mode only. We argue that this angular dependence arises naturally by considering the Dzyaloshinskii-Moriya interaction. In Ba$_2$CuGe$_2$O$_7$, the Dzyaloshinskii-Moriya interaction between two copper ions is mediated by a GeO$_4$ tetrahedron. We denote $\boldsymbol{\delta}^\perp$ as the vector that connects the center of the copper bond to the center of the GeO$_4$ tetrahedron [Fig. \hyperref[fig6]{\ref*{fig6}(a)}]. The direction of the Dzyaloshinskii-Moriya vector $\mathbf{D}$ for the bond $\boldsymbol{\delta}$ is fixed by the vectorial product $\mathbf{D} \propto \boldsymbol{\delta} \times \boldsymbol{\delta}^\perp$. For the bonds $+\boldsymbol{\delta}_x$ and $-\boldsymbol{\delta}_x$, because the GeO$_4$ center is below the copper plane, an equal component of $\mathbf{D}$ is generated in the $\mathbf{y}$ direction for both bonds. However, because of the alternating position of the GeO$_4$ center along the $\mathbf{y}$ direction (Fig. \hyperref[fig5]{\ref*{fig5}}) a sign alternating component is generated along $\mathbf{z}$. The sign-alternating component along $z$ couples $S_0^xS_\delta^y-S_0^yS_\delta^z$, and is thus zero for the cycloidal phase with the magnetic moment in the $(x,z)$ plane. This is why it has been neglected in the microscopic model of Eq. \hyperref[eq2]{(\ref*{eq2})} \cite{zheludev1999magnetic}. 

Upon applying a transverse sound wave to Ba$_2$CuGe$_2$O$_7$, we introduce the superposition of a symmetric strain and a microscopic rotation. In the theory of elasticity with a local displacement $Q_i^\mu$, the local deformation at the point $\mathbf{R}_i$ is written as $T^{\mu \nu}(\mathbf{R}_i) = \epsilon^{\mu \nu}(\mathbf{R}_i) + \omega^{\mu \nu}(\mathbf{R}_i)$, where $\epsilon^{\mu \nu}(\mathbf{R}_i) = \partial Q_i^\mu/\partial R^\nu + \partial Q_i^\nu/\partial R^\mu$ is the symmetrized microscopic strain and $\omega^{\mu \nu}(\mathbf{R}_i) = \partial Q_i^\mu/\partial R^\nu - \partial Q_i^\nu/\partial R^\mu$ the microscopic rotation \cite{luthi2007physical}. While the effect of the microscopic rotation is usually neglected, it has been shown to produce significant effects under magnetic field, introduced in Ref. \cite{lines1979elastic} as "rotational interactions". In our case, we do not use the symmetrized strains $\epsilon^{\mu \nu}(\mathbf{R}_i)$, but rather the microscopic displacement $Q_i^\mu$ directly. It allows us to explore rotational interactions as well, depending on the assumption we make for the derivative of the exchange parameters $\partial^2 \bar{J}_{\delta}^{\alpha\beta}/\partial R^\mu\partial R^\nu$. In particular, we remark that in the case of an isotropic exchange as in Eq. \hyperref[eq8]{(\ref*{eq8})}, the rotational interaction does not contribute, since only the variation of the bond distance produces a non vanishing $\partial^2 \bar{J}_{\delta}^{\alpha\beta}/\partial R^\mu\partial R^\nu$.

As shown in Fig. \hyperref[fig6]{\ref*{fig6}(b)}, atomic displacements for a transverse acoustic mode with propagation $\mathbf{k}\parallel [110]$ generates a rotation of the bond. Since the Dzyaloshinskii-Moriya vector is defined from of the vectorial product $\mathbf{D} \propto \boldsymbol{\delta} \times \boldsymbol{\delta}^\perp$, it has to be perpendicular to the rotated bond. We thus obtain that since the sound wave with in-plane polarization ($\mathbf{k} \parallel [110]$ $\mathbf{u} \parallel [1\bar{1}0]$) induces a rotation around the $z$ axis, the $y$ component of $\mathbf{D}$ varies while the bond length $||\boldsymbol{\delta}||=\delta$ remains constant, which can lead to a novel type of rotational interaction, even at zero field. In contrast, the $\mathbf{u}\parallel [001]$ polarization induces a rotation around the $y$ axis which affects the $D^z$ component perpendicular to the copper plane. Since only the $D^y$ component affects the cycloid order, we obtain that only the in-plane plane transverse acoustic mode couples to the cycloid order at zero field. This is reproduced by the following ansatz: $\bar{J}_{x}^{xz} = D_{x}^y = D \cos(Q_\delta^y/\delta)$. With this ansatz, and using $\mathbf{S}_0 = (0,0,S)$, and $\mathbf{S}_{x} = -(S\sin\alpha,0,S\cos\alpha)$ with $\alpha = \arctan(D/J)$ for the cycloidal phase of the microscopic model of Eq. \hyperref[eq2]{(\ref*{eq2})} \cite{zheludev1999magnetic}, we obtain:
\begin{equation}\label{eq9}
    \begin{split}
        \left(i^2 \frac{\partial^2 \bar{J}_{x}^{xz} }{\partial R^y\partial R^y}\right) &= \frac{D}{\delta^2}, \ \frac{\partial \bar{J}_{x}^{xz} }{\partial R^x} = \frac{\partial \bar{J}_{x}^{xz} }{\partial R^z} = 0 \\
        \Rightarrow \left(\frac{\Delta v}{v}\right)' &=  \frac{2DS^2\sin\alpha}{M_0 v^2\delta^2} \left(\mathbf{u}_k\cdot \mathbf{y}\right),
    \end{split}{}
\end{equation}{}
which is thus non-zero only for the in-plane transverse mode $\mathbf{u}_k \parallel \mathbf{y}$. 

\section{Conclusion}

In this work, we have investigated the magnetoelastic properties of Ba$_2$CuGe$_2$O$_7$ at low temperatures, and for magnetic fields applied along [001]. We focused on the elastic modes with propagation along [110], which is in the direction of nearest-neighbor copper ions. Above 1.5 K, our data match with the phase diagram obtained previously from specific-heat, magnetization, and neutron diffraction measurements \cite{muhlbauer2012phase}. We observed ultrasound anomalies consistent with three different phases: the cycloidal phase at low field, the AF-cone phase at intermediate fields, and the collinear phase at high fields. Moreover, our study reveals a specific symmetry character of the spin-lattice coupling for the different phases. The in-plane transverse mode couples predominantly to the cycloidal phase at low field, while the transverse acoustic mode $\mathbf{u}\parallel [001]$ couples stronger to the AF-cone phase. All the elastic modes couple to the collinear phase. The fact that the longitudinal mode couples only to the high-field collinear phase and, in particular, is not very sensitive to the Néel transition at zero field contradicts the exchange-striction coupling scenario \cite{luthi2007physical}. We introduced a microscopic toy model that reproduces this peculiarity by considering the variation of Dzyaloshinskii-Moriya interactions with the rotation of the copper-bond. This proposition asks for further analysis of magnetoelastic interactions in Dzyaloshinskii-Moriya magnets.

\section{Aknowledgments}
We thank K. Yu. Povarov for enlightning discussions. We acknowledge the support of the High Magnetic Field Laboratory (HLD) at Helmholtz-Zentrum Dresden-Rossendorf (HZDR), a member of the European Magnetic Field Laboratory (EMFL), the Deutsche Forschungsgemeinschaft (DFG) through SFB 1143, and the Würzburg-Dresden Cluster of Excellence on Complexity and Topology in Quantum Matter-ct.qmat (EXC 2147, Project No. 390858490).

% Stronger coupling to transverse mode rare, eg CdCr$_2$O$_4$ \cite{zherlitsyn2015novel}. It has been proposed if Ref. that in the presence of indirect exchange interactions between magnetic ion, the transverse mode will induce a the variation of angle in the exchange path, leading to a different magnetoelastic coupling than for the longitudinal mode Sr$_3$Cr$_2$O$_8$\cite{wang2016field}.

\bibliography{Biblio/biblio_JSourd_Ba2CuGe2O7.bib}

\end{document}